\documentclass[english,twocolumn]{article}
\usepackage{amsmath}
\usepackage{amssymb}
\usepackage{cite}
\usepackage[pdftex]{hyperref}
\usepackage{graphicx}
\usepackage{subfigure}
\usepackage{float}
\usepackage{subfloat}
\usepackage{makecell}
\usepackage{pbox}

\makeatletter

\usepackage{babel}

\begin{document}
	\begin{titlepage}
		\thispagestyle{empty}
		
		\bigskip
		
		\begin{center}
			\noindent{\Large \textbf
				{Universal Mass Scale for  Bosonic Fields in Multi-Brane Worlds}}\\
			
			\vspace{0,5cm}
			
			\noindent{R.I. de Oliveira Junior${}^{a}$\footnote{e-mail: ivan@fisica.ufc.br} and G. Alencar${}^{a}$\footnote{e-mail:geova@fisica.ufc.br}} 
			
			\vspace{0,5cm}
			
			{\it ${}^a$Departamento de F\'{\i}sica, Universidade Federal do Cear\'{a}-
				Caixa Postal 6030, Campus do Pici, 60455-760, Fortaleza, Cear\'{a}, Brazil. 
				
				\vspace{0.2cm}
			}

		\end{center}
		
		\vspace{0.3cm}

		\begin{abstract}
	 In this paper we find an universal mass scale for all $q-$ forms in multi-brane worlds model. It is known that this model provides an ultralight mode for the fields. However, to get this, the Lagrangians considered in the literature are not covariant. In order to solve this, we propose a covariant version to multi-localize $q-$ form fields. As a consequence of the covariance, we show that all the $q$-form fields have an ultralight mode with the same mass as the gravitational one. That way we show that there is an universal mass scale for the ultralight modes of the bosonic fields. This suggests that a new physics must emerge, for all these fields, at the same scale.

		\end{abstract}
	\end{titlepage}
	
\section{Introduction}

\label{sec:intro}
In 1998, as an extension of the Kaluza-Klein idea, Merab Gogberashvili created a shell model universe \cite{Gogberashvili:1998vx} where he deals with the hierarchy problem in a cosmological way. 
Some time ago, Randall-Sundrum (RS) presented two models in which gravity is trapped (or localized) in the membrane\cite{Randall:1999ee,Randall:1999vf}. However, unlike gravity and the scalar field, the vector field is not trapped in the brane, this became a drawback in the RS model \cite{Kehagias:2000au,Chumbes:2011zt,Zhao:2014gka,Bajc:1999mh,Ghoroku:2001zu,Oda:2000dd,Junior:2019gic,Mendes:2017hmv,Bazeia:2007nd,Fonseca:2012bw}. Despite of this, the massive modes can have peaks of probability over the brane, and show up as resonances. This tell us about unstable modes that in principle could be measured at the brane \cite{Bazeia:2007nd,Fonseca:2012bw,Cunha:2011yk,Almeida:2009jc,Liu:2009ve,Liu:2009mga,Cruz:2009ne,Zhao:2010mk,Li:2010dy,Cruz:2011kj,Landim:2011ki,Landim:2011ts,Alencar:2012en}. 

        In order to circumvent the problem of localization, some authors introduced a Dilaton coupling\cite{Kehagias:2000au}. Most of these models introduce other fields or nonlinearities to the gauge field\cite{Chumbes:2011zt}. Some years ago, Ghoroku et al proposed a mechanism that does not include new degrees of freedom and traps the gauge field to the membrane\cite{Ghoroku:2001zu}. This is based on the addition of two mass terms, one in the bulk and another on the brane $(M^{2} + c\delta(z))A_{M}A^{M}$. Despite of working, the mechanism is not covariant under a general transformation of coordinates. Beyond this, it has the undesirable feature of possessing two free parameters. In order to solve the above issues, some of the present authors found that the above term can be obtained from a bulk action $\lambda_1 R(x)A_{M}A^{M}$, where $R$ is the Ricci scalar\cite{Alencar:2014moa, Zhao:2014iqa}. Beyond solving the problem of covariance, it also eliminates from the beginning one of the free parameters. The last one is fixed by the boundary conditions, leaving no free parameters in the model.  The mechanism also keeps the advantage of not adding any new degrees of freedom. Soon later, many developments of the idea were put forward \cite{Jardim:2014cya,Jardim:2014xla,Alencar:2015awa,Alencar:2015rtc}. For example, in $D-$dimensional universe, there is the possibility of the existence of many anti-symmetric fields \cite{Alencar:2010vk,Polchinski,Polchinski2,Kulaxizi:2014yxa,Kaloper:2000xa}. However, it is a known fact that only for the $0-$form and its dual, the $(D-2)-$form, the fields are localized\cite{Duff:1980qv,Duff:2000se,Hahn:2001ze,Fu:2015cfa,Fu:2016vaj}. In order to solve this, the geometrical coupling mechanism was used in Ref. \cite{Alencar:2018cbk}. However, the coupling can be generalized to$ \lambda_1 R(x)A_{M}A^{M}+\lambda_2 R(x)_{MN}A^{M}A^{N}$ \cite{Alencar:2015oka,Freitas:2018iil}, and we gain a new undetermined parameter. However, recently, Freitas {\it et al} showed that the new parameter can be fixed by demanding consistency with Einstein Equation \cite{Freitas:2020mxr}. In fact, the last authors found that all the parameters must be fixed for any $q-$ form with the above coupling.

An interesting construction are the models with multi-gravity \cite{Kogan:1999wc,Kogan:2000xc,Kogan:2001wp}. In \cite{Kogan:1999wc}, the authors extend the RS-I model by adding a third brane with positive tension; the $ (+-+)$ model. The addition of this brane solves issues related to the cosmological constant, and also gives rise to a new phenomenology. Due to the presence of this third brane, the gravitational spectrum changes drastically. While in the RS-I the only mode that contributes to gravity is the zero one, in this model the weak gravity is a combination of the zero and first mode. This induces the idea of multi-localization of other fields, and this is done in the work \cite{Kogan:2001wp}. In this last paper the authors showed that the localization of free fields of spin $1/2,1$ and $3/2$ is not possible. As cited above, the same problem happens in the RS-I model. The way that they localize these fields is by the introduction of mass terms like $m^{2}=\alpha \sigma'(y)^{2} + \beta\sigma''(y)$, where $\sigma(y)$ is the warp factor and $\alpha,\beta$ free parameters.  This is very similar to the Ghoroku solution and shares the same problems of non-covariance and free parameters. In order to localize the zero modes of fields, the authors propose a relation between the free parameters. After this, they calculate the mass spectrum, and show that the first mode behaves like the first mode of the gravitational field as described above.  The stability and other issues about this model have been studied very recently in Refs. \cite{Lee:2021wau,Cai:2021mrw}.

 In this work we perform a complete study of bosonic fields in multi-brane worlds.  The organization of this work is as follows. In section two we review the $(+-+)$ model. Our method to multi-localize the fields in the $(+-+)$ model is done in section three. Finally, we present our conclusions and perspectives of future works.
 
\section{Review of the $(+ - +)$ Model}

The $(+-+)$ model is a generalization of the (RS-I) model where an extra brane with positive tension is added and  the negative tension brane is placed exactly in the middle of the two positive tension branes.
The metric has the usual form $ ds^{2}=e^{-2\sigma(y)}\eta_{\mu\nu}dx^{\mu}dx^{\nu} + dy^{2}$, where $\sigma(y)=k(l- ||y|-l|)$. To determine the Kaluza-Klein (KK) spectrum of gravity, we use fluctuations of the metric $\delta g_{\mu\nu}=h_{\mu\nu}(x,y)= \sum _{n=0} ^{\infty} h_{\mu\nu}^{n}(x)\psi^{n}(y)$. After a change of variables $ \frac{dz}{dy}=e^ {\sigma(y)}=g(z)= k(z_{l} - ||z|-z_{l}|) + 1$, the function $\psi^{n}(y)$ obeys the following Schrodinger differential equation

\begin{equation}\label{6}
\left( -\frac{1}{2}\partial^{2}_{z} + V(z)  \right)\hat{\psi}^{n}(z)=\frac{m_{n}^{2}}{2}e^{2\sigma(y)}\hat{\psi}^{n}(z). 
\end{equation}

That way, we have a quantum mechanical problem with delta function.
The solution for the zero mode of (\ref{6}) is found by making $m_{n}=0$. For the rest of the modes the solution is given by Bessel functions. The main characteristic of it is that the first mode has an anomalous behavior when compared with the rest of the modes. The first mode is given by
\begin{equation}\label{massall}
m_{1}=2\sqrt{2}ke^{-2kl}.
\end{equation} 
The way to obtain this first mode is given in Appendix (\ref{appendix}). The spectrum of mass for the rest of the modes is 
\begin{eqnarray}\label{11}
m_{n+1}&=&\xi_{n}ke^{-kl}, \\ \nonumber \Delta m&=&\frac{\pi}{2}ke^{-kl} ~~~~ n=1,2,3...
\end{eqnarray}
Where $\xi_{n}$ are the zeros of the function $J_{2}(\frac{m_{n}}{k}e^{kl})$. Due to the exponential factor, we see that the mass for the first mode is much smaller than the mass of the rest of the modes. 
As the first mode has a very small mass, it can contribute to gravity. It configures the bi-gravity models, where the gravity would be a combination of the zero and the first mass modes. 

The behavior for other fields assembles gravity. For a real massive scalar field in five dimensions the action is given by \cite{Kogan:2001wp}.
\begin{equation}\label{13}
S=\frac{1}{2} \int d^{4}x \int dy \sqrt{G}(G^{AB}\partial_{A}\Phi \partial_{B}\Phi + m^{2}_{\Phi}\Phi^{2}).
\end{equation}
The authors choose a mass term of the form $m_{\Phi}^{2}= \alpha(\sigma'(y)^{2}) + \beta\sigma''(y)$, where $\alpha,\beta$ are free parameters.  As we pointed before, this way of choosing the mass term is not covariant. Next, by using the standard (KK) $\Phi(x,y)= \sum_{n}\phi_{n}(x)f_{n}(y)$, and the changes of variables from the $y$ coordinate to the $z$ used above, they put the motion equation in a Schrodinger form with potential given by
\begin{eqnarray}\label{scalargeral}
V(z)=\left( \frac{9}{4} + \alpha + \beta \right) (A' (z))^ {2} + \\\nonumber \left( \frac{3}{2} -\beta \right)A''(z).
\end{eqnarray}
In order to have a zero mode localized, they found  the relation 
\begin{equation}\label{relationscalar}
\alpha=\beta^2-4\beta. 
\end{equation}
For $\beta>1$ the field is localized in the positive tension brane and $\beta<1$ in the negative one. 
Solving the Schrodinger equation for the massive modes they found for the first mode 
\begin{equation}\label{18}
m_{1}\approx \sqrt{4\nu^{2} -1}ke^{-(\nu + 1/2)kl},
\end{equation}
and for the rest of the modes 
\begin{eqnarray}\label{19}
m_{n+1}\approx \xi_{n} ke^{-kl}, \\\nonumber \; n=1,2,3,..., 
\end{eqnarray}
where $\xi_{n}$ are the roots of $J_{\nu + \frac{1}{2}}(x) $, and $\nu=\frac{3}{2}-\beta$. We should point that the light mode does not exists when 
\begin{equation}\label{betascalar}
-\frac{1}{2} \leq\nu\leq \frac{1}{2}\to 1 \leq\beta\leq 2,
\end{equation}
These parameters are get by analyzing the square root in equation (\ref{18}). To have a real mass, it is necessary that $ -\frac{1}{2} \leq\nu\leq \frac{1}{2} $. When we place it in $ \nu=\frac{3}{2}- \beta $, we find $1 \leq\beta\leq 2 $.

Let us now review the gauge field. The authors in \cite{Kogan:2001wp} used the following action
\begin{eqnarray}\label{20}
S= -\int d^{4}x \int{dy} \left[ \frac{1}{4} F_{MN} F^{MN} + \right. \\ \nonumber \left. \frac{1}{2}M A_{\mu}A^{\mu}  \right],
\end{eqnarray}
where $ F_{MN}=\partial_{M}A_{N} -\partial_{N}A_{M}$ and $M=(\beta \sigma'' (y)+\alpha(\sigma'(y))^{2})$.
Once again, with the KK decomposition $ A^{\mu}(x,y)= A^{\mu}(x)f_{n}(y) $ they obtain a Schrodinger like equation with potential given by 
\begin{eqnarray}\label{generalgauge}
V(z)=\left( \frac{1}{4} + \alpha + \beta \right) (A' (z))^ {2} + \nonumber \\ \left( \frac{1}{2} - \beta \right)A''(z)  .
\end{eqnarray}
To obtain a localized zero mode the parameter $\beta$ obeys the relation 
\begin{equation}\label{relationgauge}
\alpha=\beta^2-2\beta.
\end{equation} 
For $\beta>0$ the field is localized in the positive tension brane and $\beta<0$ in the negative one.
The expression for the first KK mode and for the rest of the tower is identical to the expression (\ref{18}), but now with 
\begin{eqnarray}\label{nugauge}
\nu= \frac{1}{2}-\beta, \nonumber\\ 0 \leq\beta\leq 1.
\end{eqnarray}
Therefore, just as in the scalar field, we get the forbidden range for $\beta$. This will be important latter. As can be seem above; for the gravity, scalar and gauge fields, we can obtain the ultra-light mode just by changing $\nu$ in expression (\ref{18}). Therefore,  in the appendix (\ref{appendix}), we consider a general potential and find, in Eq. (\ref{125}), a general expression for the first mode. This expression is valid for both multi-localization process that will be considered by us in the next sections.

\section{Covariant Multi-Localization of Bosonic Fields}
As said in the introduction, the mechanisms used by the authors of Ref. \cite{Kogan:2001wp}  has problems of non-covariance and free parameters. In this section we present two alternatives for the multi- localization of the $q-$form field: a) the geometrical coupling and b) the coupling to the dilaton field. In both methods, we will use a metric of the form $ds^{2}= e^{2A(z)}g_{MN}dx^{M}dx^{N} $.

\subsection{Multi-Localization With Geometric Quantities}
The method of localization with  geometric quantities was firstly shown in \cite{Alencar:2014moa,Zhao:2014iqa} in the context of the RS-II. As shown below, in this method the mass terms introduced are covariant, differently from \cite{Kogan:2001wp}.

\subsubsection{The Scalar Field}
For  the scalar field, the only possible covariant term to be added to the action (\ref{13}) is $\lambda R(x) \Phi^{2}$,
where $\lambda$ is the only free parameter and $R$ is the Ricci scalar. With this we see that we need to reduce the number of free parameters in order to keep general covariance. By choosing a (KK) decomposition in the form $\Phi(x,z)=e^ {-\frac{3}{2}A}\phi(x)\psi(z) $, and placing the Ricci scalar in the $z$ coordinate: $R=-e^ {-2A}(8A" +12(A')^{2})$, we finally arrive in the Schrodinger equation with potential given by
 \begin{eqnarray}\label{c}
V(z)&=&c_1 A''   +c_2 A'^{2},\nonumber\\ c_1&=&\left( \frac{3}{2} -8\lambda\right)\nonumber\\,c_2&=&\left( \frac{9}{4} - 12\lambda\right).
\end{eqnarray}
For the massless case it is easy to see that the solution must be given by $e^{c_1A}$, with $c_1^{2}=c_2$. With this we find two solutions $ \lambda=3/16,\lambda=0$. Lets us compare this with the solution found in Ref \cite{Kogan:2001wp}. Comparing the potential (\ref{scalargeral}) with (\ref{c}),  we get $\alpha=-20\lambda,\beta=8\lambda$. If $ \lambda=\frac{3}{16}$, we have $\beta=\frac{3}{2}$. If $\lambda=0$, we have $\beta=0$. Now, the integral in the extra dimension is give by $\int e^ {2c_{1}A}dz$ and in order to have the field localized in the positive branes, we need $2c_{1}>1$. As $c_{1}$ is not a free parameter, it is $c_{1}=\frac{3}{2}$, the zero mode is localized only in such branes. 

Considering now $\lambda=\frac{3}{16}$, we re obtain the relation (\ref{relationscalar}), used by the authors of Ref. \cite{Kogan:2001wp} to ensure localization. Also,  In Ref. \cite{Kogan:2001wp} the authors find that  for $\beta<1$ the scalar field is localized in the positive tension brane and for $\beta>1$ in the negative one \footnote{In fact the authors committed a small mistake and inverted the range.}. However now $\beta$ is not a free parameter and the only  allowed value is $\beta=3/2$. Therefore, imposing general convariance will provides a radically different behavior for the model. First of all,  the scalar field must be localized only in the negative tension brane. Second, and more important, $\beta=3/2$ implies that it is in the forbidden range of Eq. (\ref{betascalar}). Therefore the ultralight mode does not exist and this will gives a very different phenomenological consequences to the model. If the ultralight mode does not exist, we can not ensure the localization of the zero mode. This fact forces us to choose the value $\lambda=0$. This choice leads to $\beta=0$, and now there is an ultralight mode and also the zero mode is localized in the positive tension branes, not in the negative one as would be if we chose $\lambda=\frac{3}{16}$. Then, we see that the scalar field does not needs a mass term in order to be localized. Also, the ultralight mode has a mass identical to the gravitational field. The rest of the spectrum is given in (\ref{19}),with $\nu=\frac{3}{2}$.

\subsubsection{The Gauge Field}

We now focus on the gauge field. The action proposed by \cite{Kogan:2001wp}, equation (\ref{20}) is not invariant under a general transformation of coordinates. In order to correct such a problem we propose a path similar to the one of the scalar field and add $-\frac{\lambda}{2}R A_{M}A^{M}$ to the action \label{20}. This action has been considered in Ref. \cite{Alencar:2014moa}  in the context of the RS-II model. After performing the separation of variables $ A^ {\nu}(x,z)= e^ {-\frac{A}{2}}A^ {\nu}(x)\psi(z)$ the authors found a Schrodinger equation with potential
\begin{eqnarray}\label{i}
V&=&c_1 A''   +c_2 A'^{2},\nonumber\\,c_1&=& \frac{1}{2} -8\lambda,\nonumber\\,c_2&=& \frac{1}{4} -12\lambda.
\end{eqnarray}
The solution to the massless case is therefore again $\psi=\propto e^{c_1A}$, with $c_1^2=c_2$. With this we get
$\lambda=-1/16;c_1=1$. With this solution we get that the integral in the extra dimension is $\int e^{2A(z)}dz$ and the field is localized in the positive tension brane. 

Let us now compare this with the results of Ref.\cite{Kogan:2001wp}. First we compare our potential with (\ref{generalgauge}) to find $\beta=8\lambda,\,\alpha=-20\lambda$. With this we see that (\ref{relationgauge})  is automatically satisfied.   This condition was used by the authors of Ref. \cite{Kogan:2001wp} to ensure localization.  Also,  In Ref. \cite{Kogan:2001wp} the authors find that the Gauge field is localized in the positive tension brane for $\beta<0$  and in the negative one for $\beta>0$\footnote{Again we have corrected the range.}. However now $\beta$ is not a free parameter and the only  allowed value is $\beta=-1/2$. Therefore, imposing general covariance will provides a specific behavior for the model. First of all,  the Gauge field must be localized only in the positive tension branes. Second, and more important, $\beta=-1/2$ is out of the forbidden range  (\ref{nugauge}). Therefore the ultralight mode exists, and is fixed. The rest of the mass, again, will be given by (\ref{19}) with 
$\nu=1$.

Now we can generalize the above model by considering a more general coupling given by $\lambda_{1}R A_{M}A^{M} - \lambda_{2}R_{MN}A^{M}A^{N}$.
This action was used in the context of RS-II in Ref. \cite{Freitas:2018iil}. The process to obtain the Schrodinger equation is similar to the one for just the Ricci scalar, and we get  
\begin{eqnarray}\label{i2}
c_1&=& \frac{1}{2} -8\lambda_{1} -\lambda_{2},\nonumber\\,c_2&=& \frac{1}{4} -12\lambda_{1} -3\lambda_{2}.
\end{eqnarray}
The solution to the zero mode is obtained with $c_1^2=c_2$, and now we get
\begin{eqnarray}\label{condgauge2}
\lambda_2^{\pm}&=&-(1+8\lambda_{1})\pm\sqrt{12\lambda_1+1};\nonumber\\c_1^{\pm}&=&\frac{3}{2}\mp\sqrt{12\lambda_1+ 1} 
\end{eqnarray}
Of course, we must have $\lambda_1\ge-1/12$ in order to have a real root. This leads to an extra dimension integral $\int e^{2c_1A(z)}dz$ and the field is localized in the positive tension brane if $2c_1>1$ and in the negative one if $2c_1<0$.  With this we get that  the range of allowed values is given by
\begin{eqnarray}\label{range}
-\frac{1}{12}\le\lambda_{1} <0,\nonumber\\-2<\lambda_{2} \leq -\frac{1}{3}.
\end{eqnarray}

Once more we compare our potential  with (\ref{generalgauge}). The comparison leads to
$\beta=(8\lambda_{1}+\lambda_{2}),\, \alpha=-(20\lambda_{1}+4\lambda_{2})$. As expected,  the above expressions imply that the relation (\ref{relationgauge})  is automatically satisfied. We also obtain, from Eq.  (\ref{range}), the range of values $-8/3\leq\beta<0.$ The above values are out of the forbidden range (\ref{nugauge}) and the ultralight mode exists. We should point the interesting particular case $\lambda_{1}=0$. With this we get a coupling just with the Ricci tensor and $\beta=\lambda_{2},\alpha=-4\lambda_{2} $. With the condition $c_{1}^{2}=c_{2}$, we get $\lambda_{2}=-2$, then $\beta=-2$ and $\alpha=8$. Therefore we get a fixed ultralight mode.

However, for the general case, we can not  completely fix the parameters $\lambda_{1}$ and $\lambda_{2}$. It is necessary something else. As said in the introduction, we can impose the further condition that the localization of the field is consistent with EE in the limit of large radius. Using this, Freitas {\it et al}  found that we must fix  $\lambda_{1}=-1/12,\,\lambda_{2}=-1/3$\cite{Freitas:2020mxr}. With these values we have $\beta=-1$, and we get a fixed ultralight mode that is identical to the gravitational field. The rest of the (KK) modes is calculated with (\ref{19}),with  $\nu=\frac{3}{2}$.

  \subsubsection{The Kalb-Ramond Field}
Here we analyze the multi-localization of the Kalb-Ramond Field. Here, from the beginning we use  both couplings. The case, in the context of just one brane, is discussed in \cite{Alencar:2018cbk}. The action is given by
\begin{eqnarray}\label{103}
S&= &\int d^{5}x \sqrt{-g} \Bigg[ \frac{1}{12} F_{M_{1}M_{2}M_{3}} F^{M_{1}M_{2}M_{3}}\nonumber\\
&+& \frac{\lambda_{1}}{4}R A^{M_{2}N_{2}}A_{M_{2}N_{2}} \nonumber \\
&+& \frac{\lambda_{2}}{4}g^{N_{1}N_{2}}R^{M_{1}M_{2}}A_{M_{1}N_{1}}A_{M_{2}N_{2}} \Bigg].
\end{eqnarray}
Where $ F_{M_{1}M_{2}M_{3}}=3\partial_{[M_{1}}A_{M_{2}M_{3}]}$. The authors of Ref. \cite{Alencar:2018cbk} use the separation  $A^{\mu\nu}_{T}= e^{\frac{A}{2}}\tilde{A}^{\mu\nu}_{T}(x)\psi(z)$ to find a Schrodinger equation with
\begin{eqnarray}\label{kalb}
c_{1}&=&-\frac{1}{2} -8 \lambda_{1}-\lambda_{2},\nonumber\\ c_{2}&=& \frac{1}{4} - 12\lambda_{1} -3\lambda_{2}.
\end{eqnarray}
The solution to the zero mode is obtained if $c^{2}_{1}=c_{2}$ and we find 
$
	\lambda_{2}^{\pm}= -(8\lambda_{1} +2) \pm \sqrt{(12\lambda_{1}+4)}.
$
We have a localized zero mode in the positive tension brane for $2c_{1}>1$ and for $2c_{1}<0$ in the negative one. This leads to 
$
-4\le12\lambda_1<-3,-3<3\lambda_2\le2.
$
For the light mode we use Eq. (\ref{125}) of the appendix to arrive at
$
m_{1}=\sqrt{4c_1^{2}-1}ke^{-(c_1+1/2)x}.
$
Therefore, we see that the condition for localization is exactly the one to get an ultra-light mode. Again, we do not know the exact value of the parameters. However, we can use  the consistency condition developed in \cite{Freitas:2020mxr}. For the KR field the values of the parameters are fixed to $\lambda_{1}= -1/3,\lambda_{2}=2/3$.
With this value we get a localized zero mode and an ultra light mode given by $c_{1}=3/2$. Again, we get that the only consistent ultralight mode has the same mass as in the gravity case.

\subsubsection{$q-$forms}
The action for the $q-$form coupled with the geometric quantities(Ricci scalar and Ricci tensor) is \cite{Alencar:2018cbk}.
\begin{eqnarray}
S=&-& \int d^{D}x\sqrt{-g} \Bigg[ \frac{1}{2(q+1)!}F_{M_{1}...M_{q+1}}F^{M_{1}...M_{q+1}}\nonumber \\ 
&+& \frac{\lambda_{1}}{2q!} R A_{M_{2}...M_{q+1}}A^{M_{2}...M_{q+1}}\nonumber  \\
 &+& \frac{\lambda_{2}}{2q!} R^{M_{1}M_{2}} A_{M_{1}M_{3}...M_{q+1}}A_{M_{2}}^{\;\;\;\;M_{3}...M_{q+1}}  \label{110} \Bigg],
\end{eqnarray}
where $F_{M_{1}M_{2}...M_{q+1}}=(q+1) \partial_{[M_{1}}A_{M_{2}M_{3}...M_{q+1}]} $.
The authors of Ref. \cite{Alencar:2018cbk} use the separation  $A^{\mu_{1}...\mu_{q}}_{T}=\hat{A}^{\mu_{1}...\mu_{q}}_{T}(x)e^{-\frac{\alpha_{q}A}{2}}\psi(z)$ to find a Schrodinger equation with
\begin{eqnarray}\label{coeficientesg}
 c_{1}&=&\frac{\alpha_{q}}{2} -2\lambda_{1}(D-1)-\lambda_{2},\nonumber\\ c_{2}&=&\frac{\alpha^{2}_{q}}{4} - \lambda_{1}(D-1)(D-2)-\lambda_{2}(D-2)
\end{eqnarray}
where $\alpha_{q}=[D-2(q+1)]$. The solution for the zero mode is $\psi=e^ {c_{1}A}$. With that we get $c_{1}^{2}=c_{2}$. To have a $q-$form localized in the positive branes we need $2c_{1}>1$,and for the negative one $ 2c_{1}<0$. With these conditions we find the range of values for $\lambda_{1}$ and $\lambda_{2}$\cite{Freitas:2020mxr}.  To fix it we need to use the consistency conditions found in the same reference and they finally get
 \begin{eqnarray}
 \lambda_{1}&=&\frac{2\alpha_{q}(D-2)-(D-2)^{2}-\alpha_{q}^{2}}{4(D-1)(D-2)},\nonumber\\
 \lambda_{2}&=&-\frac{[\alpha_{q}(D-2)-\alpha_{q}^ {2}]}{2(D-2)}. 
 \end{eqnarray}
With these values for $\lambda_{1}$ and $\lambda_{2}$ we get $c_{1}=\frac{(D-2)}{2}$ and $c_{2}=\frac{(D-2)^{2}}{4} $. This leads to the integral over the extra dimension $\int e^{(D-2)A}dz$.
 
 One interesting fact that deserves to be mentioned is the analogy between the $q-$form as described here, and the gravitational field in $D$-dimensions. To show it we need to remember that the gravitational field is very similar to the free scalar field, in other words, to the $q=0$ form. With that in mind we have  $\alpha_{q}= D-2$. Placing it in the value of $c_{1}$ found in (\ref{coeficientesg}),and making $\lambda_{1}=\lambda_{2}=0$, we have $c_{1}=\frac{D-2}{2}$. As the solution is of the form $\psi=e^ {2c_{1}A}$, we get $\psi=e^ {(D-2)A}$, and this leads to the integral over the extra dimension $\int e^{(D-2)A}dz$. This is a very important result because it tell us that the ultralight mode of all the  $q-$form fields are identical to the gravitational one, with mass given by (\ref{massall}). If we consider $D=5$, we always will get $c_{1}=\frac{3}{2}$ for bosonic fields. Substituting this value for $c_{1}$ in  (\ref{massall}), we get the mass of the ultralight mode for the gravitational field (\ref{massall}).  In the next section we discuss the multi-localization with the dilaton field, and we will show that we get the same results found here.

\subsection{Multi-Localization With the Dilaton}
Now we present the localization of $q-$forms with the coupling with dilaton. The model has been studied in Ref. \cite{Landim:2010pq} and we have a dilaton field with solution $\pi=-\sqrt{3M^{3}}A(z)$.
The action for the $q-$form in a $p$-brane in $D$-dimensions is \cite{Landim:2010pq}
\begin{equation}
S=- \frac{1}{2(q+1)!}\int d^{D}x\sqrt{-g} e^{-\lambda\pi} F_ {M_{1}...M_{q+1}}F^ {M_{1}...M_{q+1}}.
\end{equation}
By performing the separation $A^ {\mu_{1}...\mu_{q}}(x,z)=e^ {-\frac{\rho_{q}A}{2}}A^ {\mu_{1}...\mu_{q}}(x)\psi(z) $ the authors of Ref\cite{Landim:2010pq} find the general potential with $c_{1}=\frac{\rho_{q}}{2}, \, c_{2}=\frac{\rho_{q}^ {2}}{4},$ where $\rho_{q}= D-2(q+1) +\lambda \sqrt{3M^ {3}} $. 

From the above solution, the integral over the extra dimension is given by $\int dz e^{2c_{1}A}$. By a simple inspection in this action, we see that the condition to be a $q$-form localized in the positive branes is $2c_{1}>1$, and in the negative brane $2c_{1}<0$. This leads, respectively, to $\lambda> \frac{2q +3-D}{\sqrt{3M^{3}}} $ and $ \lambda < \frac{2(q+1) -D}{\sqrt{3M^ {3}}}$. As we already know, the way to fix the free parameter we given in Ref. \cite{Freitas:2020mxr} and we have $ \lambda=\frac{2q}{\sqrt{3M^ {3}}}$. Once more we can compare the $q$-form with the free gravitational field in $D$-dimensions. After fixing $\lambda= \frac{2q}{\sqrt{3M^{3}}}$, we have $ \rho_{q}=(D-2)$, then $2c_{1}=(D-2)$ just like the gravity case in $D$ dimensions.  This means that the mass of the ultralight mode of the $q$-form coupled to dilaton, also is given by (\ref{massall}).

\section{Conclusion}
Multi-brane worlds were introduced some time ago by Kogan et al. They discovered that the introduction of a third brane in the RSI model modifies gravity at large scales \cite{Kogan:2000xc}.  The authors found an ultralight mass modes for all the fields, given by $m_{1}^2=m_{1}\approx \sqrt{4\nu^{2} -1}ke^{-(\nu + 1/2)kl}$,
where $\nu=3/2$ for gravity and has a range of possibilities for all the other fields. This would lead to new physics and the model was named``Multi-Brane worlds". In order to reach the localization for fields other than gravity, the authors introduced  a bulk and a brane mass terms given by $
m^{2}= \alpha(\sigma'(y)^{2}) + \beta\sigma''(y)$, where $\alpha,\beta$ are free parameters. Therefore the mass of such modes depends on a free parameter and are not fixed. Another point is that, despite working, the above mass term breaks the general covariance of the system.

In order to solve the above problems we used a geometrical coupling proposed by some of us in Ref. \cite{Alencar:2014moa}. For this we first consider the scalar and vector fields. The key point to note  is that  $R=-(8\sigma' +20(\sigma')^{2}),$ which is very similar to the above mass term. The covariance problem, in this way, is solved by introducing a mass term given by $\gamma R\Phi^2$. Note that with this we reduced the number of free parameters to one. Therefore covariance eliminates one free parameter. Next, we apply consistency conditions found by some of us in Ref. \cite{Freitas:2020mxr}. This fixes all the parameters and we are left with light modes given by $m_{1}^2=2\sqrt{2}ke^{-2kl}$. Beyond being fixed, this mass is exactly the same as the ultralight mode of gravity.

Next, we generalize the same mechanism to obtain multi-localization of $q-$forms.We do it by imposing general covariance  through a general coupling with $R$ and $R_{MN}$. Then, with these  consistency conditions we fix all the parameters of the model.  We compute the mass spectrum, and after imposing  consistency conditions we find that the ultralight modes are all the same as the gravity one. This is very curious, and in order to test if this is a characteristic of geometrical couplings, we also consider the localization by using the dilaton field. Again we find the same ultralight mode for all the bosonic fields. Therefore, we conclude that; for all the fields and localization mechanisms considered here,  the ultralight mode is identical to the gravitational one. In this way we established an universal mass scale for the bosonic fields. We have the Physics due to the zero mode, and then the next scale of energy is valid for all the first modes of all the bosonic fields. A good generalization would be to consider fields of spin $1/2$ and $3/2$. This will be the subject of our next studies.

\section*{Acknowledgments}

The authors would like to thanks Alexandra Elbakyan and sci-hub, for removing all barriers
in the way of science, and to Makarius Tahim for useful conversations. We acknowledge the financial support provided by the Conselho
Nacional de Desenvolvimento Cient\'ifico e Tecnol\'ogico (CNPq) and Funda\c{c}ao Cearense de
Apoio ao Desenvolvimento Cient\'ifico e Tecnol\'ogico (FUNCAP) through PRONEM PNE0112- 00085.01.00/16

\appendix

\section{Mass Spectrum in the  $(+-+)$ Model  } \label{appendix}
In this appendix we show how tho get the spectrum for the fields in the  $(+-+)$ model. All the fields follow a Schrodinger like equation with 
\begin{eqnarray}
V(z)= \frac{(c_{1} + c_{1}^ {2})k^{2}}{g(z)^{2}} -\nonumber\\ \frac{2c_{1}k (\delta(z) + \delta(z-2l) -\delta(z-l))}{g(z)}
\end{eqnarray}
The solution for the massive modes is of the form
\begin{eqnarray}
\psi^{n}= \sqrt{\frac{g(z)}{k}}\left[ C_1 J_{c_{1}+\frac{1}{2}}\left( \frac{m_{n}}{k}g(z)\right)  + \right. \\ \nonumber \left. C_2  Y_{c_{1}+\frac{1}{2}}\left( \frac{m_{n}}{k}g(z)\right)\right].\label{121}
\end{eqnarray}
Using the boundary conditions for the wave function and its first derivative we arrive at the condition
\begin{eqnarray}
\left[J_{c_{1}-1/2}(a)Y_{c_{1}-1/2}(ax) -  Y_{c_{1}-1/2}(a)J_{c_{1}-1/2}(ax)\right]\times\\ \nonumber \left[Y_{c_{1}+1/2}(xa)J_{c_{1}-1/2}(a)  - J_{c_{1}+1/2}(ax)Y_{c_{1}-1/2}(a)\right]=0.
\end{eqnarray}
Where $x=g(z)$ and $a=\frac{m_{n}}{k}$.
We then expand this expression for small mass on Mathematica, to find
\begin{equation}\label{125}
m_{1}=\sqrt{4c_1^2-1}ke^{-(c_1+1/2)x}.
\end{equation}
With the second expression, and considering that $J(x)$ is very small compared with $Y(x)$ for small arguments, we have
\begin{equation}
J_{c_{1}+1/2}(a*x)Y_{c_{1}-1/2}(a)=0\to m_{n}=\xi k e^{-kl},
\end{equation}
where $\xi$ are the roots of the Bessel functions $J_{c_{1}+1/2}(a*x)$.


\end{document}